\documentclass[aps,prl,reprint,showpacs]{revtex4-1}%
\usepackage[dvips]{graphicx}%
\usepackage{bm}% 
\usepackage{amsmath}
\usepackage[colorlinks=true,linkcolor=blue,citecolor=red ,urlcolor=blue]{hyperref}

\begin{document}

\title{Quantum phase transitions of three-level atoms interacting with a
one-mode electromagnetic field}

\author{S. Cordero}
\email{sergio.cordero@nucleares.unam.mx}
\author{R. L\'opez--Pe\~na}
\email{lopez@nucleares.unam.mx}
\author{O. Casta\~nos}
\email{ocasta@nucleares.unam.mx}
\author{E. Nahmad--Achar}
\email{nahmad@nucleares.unam.mx}
\affiliation{%
Instituto de Ciencias Nucleares, Universidad Nacional Aut\'onoma de
M\'exico, Apartado Postal 70-543, 04510 M\'exico DF, M\'exico}

%\date{}

\begin{abstract}
We apply the energy surface method to study a system of $N_{a}$
three-level atoms interacting with a one-mode radiation field in the
$\Xi$-, $\Lambda$- and $V$--configurations. We obtain an estimation
of the ground-state energy, the expectation value of the total number of excitations,
and the separatrix of the model in the interaction
parameter space, and compare the results with the exact solutions. We have
first- and second-order phase transitions, except for the
$V$--configuration which only presents second-order phase-transitions.
\end{abstract}

\pacs{42.50.Ct,42.50.Nn,73.43.Nq,03.65.Fd}%

\maketitle

%\section{Introduction}

1. {\it Introduction.} The Tavis-Cummings Model~\cite{tavis-cummings1968}, which describes the
interaction of a collection of $N$ two-level atoms with a quantized
electromagnetic field in the dipolar and rotating-wave approximations (RWA),
has an extensive use in quantum optics~\cite{manko}. Recently this model has been
physically realized using a QED cavity with Bose-Einstein
condensates~\cite{baumann2010,nagy2010}. Particularly interesting has
been the investigation of the phase transitions of the
system in the thermodynamic limit~\cite{hepp&lieb1973}, and at zero
temperature~\cite{buzek2005,castanos2009}.

The physics of three-level systems interacting with one or two
quantized modes of the electromagnetic field is very rich and many
special dynamical situations have been
studied. In particular, a formalism to describe one three-level atom interacting with 
a one- or a two-mode field has been discussed together with the atomic level 
occupation probabilities, coherence properties, photon probability distribution, 
fluctuations, and the evolution of squeezing in a series of works~\cite{li1987}. 
For one three-level atom interacting with a one- or a two-mode field, it was found 
that the phase distribution properties of the field reflect the collapses and 
revivals of the level occupation probabilities. However, for the two-mode case, 
there are exceptions and the collapses and revivals are decorrelated from the 
phase field~\cite{alikskenderov1991}.  
 
A comprehensive review of
the dynamical interaction of an atom with radiation in the framework
of the Jaynes--Cummings type has been done by Yoo and
Eberly~\cite{yoo&eberly1985}. However, by means of the thermodynamic 
Green function the phase transitions of the Dicke
model, including all modes of the radiation field,  for atoms confined in 
a cubic resonance wavelength, have been considered for a 
finite temperature in~\cite{sung-bowden1979}. Recently,
there has been a semi-classical treatment of the phase transitions in a
system of three-level atoms in the $\Lambda$-configuration interacting
with a two-mode quantized electromagnetic field without the RWA, using
however the generalized Holstein-Primakoff transformation to find the
separatrix of the system in the thermodynamic
limit~\cite{brandes2011}.

%Figura 1
\begin{figure}
\begin{center}
\includegraphics[width=0.9\linewidth]{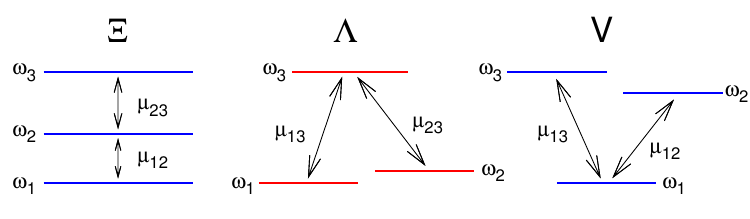}
\end{center}
\vspace{-0.5cm}
\caption{Schematic representation of the three atomic configurations. $\hbar \omega_i$ denotes the energy of the $i$-th level, and $\mu_{ij}$ the dipolar coupling between levels $i$ and $j$. We use $\hbar=1$ throughout. }\label{f1}
\end{figure}

The main contribution of this manuscript is the establishment of the separatrix of the three-level system interacting with a one-mode electromagnetic field for the three atomic configurations, $\Xi$, $\Lambda$, and $V$ (cf. Fig.~\ref{f1}). Each separatrix determines in control parameter space (dipolar strenghts) where the quantum phase transitions take place, and which are of first- and second-order depending on the values of the dipolar interactions. The presence of the quantum phase transitions can be clearly seen in the calculation of the ground state energies and in the expectation value of the total number of excitations, as functions of the dipolar couplings $\mu_{12}$, $\mu_{13}$, and $\mu_{23}$. The agreement with the corresponding exact quantum calculations is remarkable, in spite of considering a small number $N_a$ of atoms ($N_a=2$ and $N_a=10$).  Furthermore,  by taking $\mu_{23}=0$ in the $\Xi $- and $\Lambda$- configurations, and $\mu_{13}=0$ in the $V $ case, the corresponding Hamiltonian systems describe two-level problems in the manner of Tavis-Cummings, giving consistency to our results. In these cases it is straightforward to check that the separatrix coincides with the one established by Hepp and Lieb~\cite{hepp&lieb1973}. 

The organization of the paper is as follows. In Section 2 we establish
the model and present the basic formulation of
the problem. In Section 3 an analytic expression
for the energy surface of the system is obtained, and use it to obtain an
estimation of the ground state of the system in the three basic
configurations.  In Section 4 we solve the Hamiltonian for a finite
number of particles and compare the results with those previously
obtained. Finally, a summary of the general results and conclusions is
outlined in Section 5.

2. {\it Model Hamiltonian.} We consider a quantum system of $N_{a}$ three-level atoms, each atom
being able to occupy one of three levels characterized by energies
$\omega_{1}$, $\omega_{2}$ and $\omega_{3}$, interacting dipolarly
with a one--mode field of frequency $\Omega$, and
assume the rotating-wave approximation. We consider
$\omega_{1}\le\omega_{2}\le\omega_{3}$. The Hamiltonian describing
this system can be written as~\cite{yoo&eberly1985}
   \begin{eqnarray}
	&&\hat{H}=\Omega\,\hat{a}^{\dagger}\hat{a}
	+\sum^{3}_{k=1} \omega_{k}\,\hat{A}_{kk}
	 -\frac{\mu_{12}}{\sqrt{N_{a}}}\left(\hat{a}\,\hat{A}_{21}
	 +\hat{a}^{\dagger}\,\hat{A}_{12}\right)\nonumber \\ 
	-&&\frac{\mu_{13}}{\sqrt{N_{a}}} \left(\hat{a}\,\hat{A}_{31}
	 +\hat{a}^{\dagger}\,\hat{A}_{13}\right) -\frac{\mu_{23}}{\sqrt{N_{a}}} \left(\hat{a}\,\hat{A}_{32}
	 +\hat{a}^{\dagger}\,\hat{A}_{23}\right)
	\label{eq01}
   \end{eqnarray}
where the first two terms on the r.h.s. are related to the free atomic and field parts. The $\hat{a}$, $\hat{a}^{\dagger}$, 
denote the annihilation and creation operators of the radiation field, and the operators
$\hat{A}_{ij}$ are the generators of a $U(3)$ algebra. The operators
$\hat{a}$, $\hat{a}^{\dagger}$, satisfy the Heisenberg--Weyl algebra
commutators
    \begin{equation}
	\left[\hat{a},\,\hat{a}^{\dagger}\right]=\hat{1}\ ,\quad
	\left[\hat{a},\,\hat{a}\right]=0\ ,
	\nonumber
	\label{eq02}
    \end{equation}
while the atomic operators $\hat{A}_{ij}$ (weight for $i=j$, lowering for $i>j$ and raising with $i<j$) 
fulfill the $U(3)$-algebra commutation relations
    \begin{equation}
	\left[\hat{A}_{ij},\,\hat{A}_{kl}\right]=\delta_{jk}\hat{A}_{il}
	-\delta_{il}\hat{A}_{kj}\ .
	\nonumber
	\label{eq03}
    \end{equation}
The usual three-level atomic arrangements are obtained
from~(\ref{eq01}) imposing $\mu_{13}=0$ for the $\Xi$--configuration,
$\mu_{12}=0$ for the $\Lambda$--configuration, or $\mu_{23}=0$ for the
$V$--configuration. In each case there are two constants of 
motion: the total number of atoms
    \begin{equation}
	\hat{N}_{a}=\hat{A}_{11}+\hat{A}_{22}+\hat{A}_{33}\ ,
	\label{eq04}
    \end{equation}
and the total number of excitations 
    \begin{eqnarray}
	\hat{M}_{\Xi}&=&\hat{a}^{\dagger}\hat{a}+\hat{A}_{22}
	+2\,\hat{A}_{33}\ ,
	\nonumber\\
	\hat{M}_{\Lambda}&=&\hat{a}^{\dagger}\hat{a}
	+\hat{A}_{33}\ ,
	\nonumber\\
	\hat{M}_{V}&=&\hat{a}^{\dagger}\hat{a}+\hat{A}_{22}
	+\hat{A}_{33}\ .
	\label{eq07}
    \end{eqnarray}

3. {\it Energy surface.} In order to obtain an energy surface, we use as a trial state the
direct product of coherent states in each subspace.  The use of
coherent states as trial states lets us determine in analytic form the
expectation values of matter and field observables. Thus we use the
tensor product of Heisenberg-Weyl $HW(1)$ coherent states for the radiation part, $\vert \alpha \}=e^{\alpha\,\hat{a}^{\dagger}}\,\vert 0\rangle$, and $U(3)$ coherent states for
the atomic part. The unnormalized $U(3)$ coherent state can be constructed 
by taking the exponential of the lowering
generators acting on the highest weight states of $U(3)$, 
   \begin{equation}
	\big|[h_{1},h_{2},h_{3}]
        \gamma_{1},\gamma_{2},\gamma_{3}\big\}
       =e^{\gamma_{3} \hat{A}_{21}}
	\, e^{\gamma_{2} \hat{A}_{31}} \,  e^{\gamma_{1} \hat{A}_{32}} 
         \vert\,[h_{1},h_{2},h_{3}] \rangle \ , \nonumber
	\label{eq08}
   \end{equation}
where we denote by
      \begin{equation}
        \vert\,[h_{1},\,h_{2},\,h_{3}]\rangle
        \equiv\left|\begin{array}{ccccc}
	h_{1}&&h_{2}&&h_{3}\\
	&h_1&&h_2&\\
	&&h_1&&
	\end{array}\right\rangle\ ,
	\nonumber
      \end{equation}
the highest weight state of the Gelfand-Tsetlin
basis of the irreducible representation $[h_{1},\,h_{2},\,h_{3}]$
of $U(3)$~\cite{gelfand-tsetlin1950}. The  action of the
raising operators $\hat{A}_{ij}$ on the highest
weight state vanishes:
$\hat{A}_{ij}\,\vert\,[h_{1},\,h_{2},\,h_{3}] \rangle=0$.

In this contribution we only consider the completely symmetric representation
$[N_{a},\,0,\,0]$ of $U(3)$, where $N_{a}$ denotes the number of
atoms. Therefore, the trial state is the tensor product
      \begin{equation}
	\vert N_{a},\alpha,\gamma_{2},\gamma_{3}\}
	=e^{\alpha\,\hat{a}^{\dagger}}\vert 0\rangle \otimes
	\,e^{\gamma_{3}\hat{A}_{21}}
	\,e^{\gamma_{2}\hat{A}_{31}}
        \vert [N_{a},0,0] \rangle\ .
	\nonumber
      \end{equation}
The parameter $\gamma_{1}$ does not appear in the coherent state
because $\hat{A}_{32}\,\vert\,[N_{a},\,0,\,0] \rangle=0$.

The expectation value with respect to this state of the model
Hamiltonian~(\ref{eq01}) gives the energy surface
    \begin{eqnarray}
	&&{\cal H}(\varrho,\varrho_{2},\varrho_{3},\vartheta_{1},
	 \vartheta_{2},\vartheta_{3})=\Omega\,\varrho^{2}
	+N_{a}\frac{\omega_{1}+\omega_{2}\,\varrho_{3}^{2}
	+\omega_{3}\,\varrho_{2}^{2}}{1+\varrho_{2}^{2}
	+\varrho_{3}^{2}}\nonumber\\
	&-&2\sqrt{N_{a}} \varrho\frac{\mu_{12} \varrho_{3}
         \cos\vartheta_{3}
	+\mu_{13} \varrho_{2} \cos\vartheta_{2}
	+\mu_{23} \varrho_{2} \varrho_{3} \cos\vartheta_{1}}{1
	+\varrho_{2}^{2}+\varrho_{3}^{2}}\ ,\nonumber
	 \label{eq12}
     \end{eqnarray}
where we have used the identifications
$\gamma_{k}=\varrho_{k}\,\exp(i\,\varphi_{k})$, with $k=2,\,3$, 
and $\alpha=\varrho\,\exp(i\,\varphi)$. Additionally, 
the angles $\vartheta_{1}=\varphi-\varphi_{2}+\varphi_{3}$,
$\vartheta_{2}=\varphi-\varphi_{2}$, and
$\vartheta_{3}=\varphi-\varphi_{3}$ were defined.
    
The minima for the energy surface are obtained when $\vartheta_{i}=0,\,\pi$ for $i=1,2,3$, and 
    \begin{equation}
	\mu_{12}\,\cos\vartheta_{3}>0\,,
         \ \mu_{13}\,\cos\vartheta_{2}>0\,,
	 \ \mu_{23}\,\cos\vartheta_{1}>0\ .
	 \nonumber
    \end{equation}
To have an intensive quantity we divide the energy surface ${\cal H}(\varrho,\varrho_{2},\varrho_{3},\vartheta_{1},
\vartheta_{2},\vartheta_{3})$ by $N_{a}$, and use the conditions of a minimum for
the angles $\vartheta_{i}$ to obtain
     \begin{eqnarray}
	&&{\cal E}(\bar{\varrho},\,\varrho_{2},\,\varrho_{3})=
	\,\bar{\varrho}^{2}
	+\frac{\omega_{1}+\omega_{2}\,\varrho_{3}^{2}
	+\omega_{3}\,\varrho_{2}^{2}}{1+\varrho_{2}^{2}
	+\varrho_{3}^{2}}\nonumber\\
	&&\quad-2\,\bar{\varrho}\,\frac{
	\left|\mu_{12}\right|\,\varrho_{3}
	+\left|\mu_{13}\right|\,\varrho_{2}
	+\left|\mu_{23}\right|\,\varrho_{2}\,\varrho_{3}}{1
	+\varrho_{2}^{2}+\varrho_{3}^{2}}\ ,
	\label{eq14}
     \end{eqnarray}
where $\bar{\rho}=\rho/\sqrt{N_a}$, and $\omega_{k}$ and $\mu_{kl}$ are now measured in units of
$\Omega$. In what follows we will set $\Omega=1$.

% Figure 2
\begin{figure} \begin{center}
\includegraphics[width=0.45\linewidth]{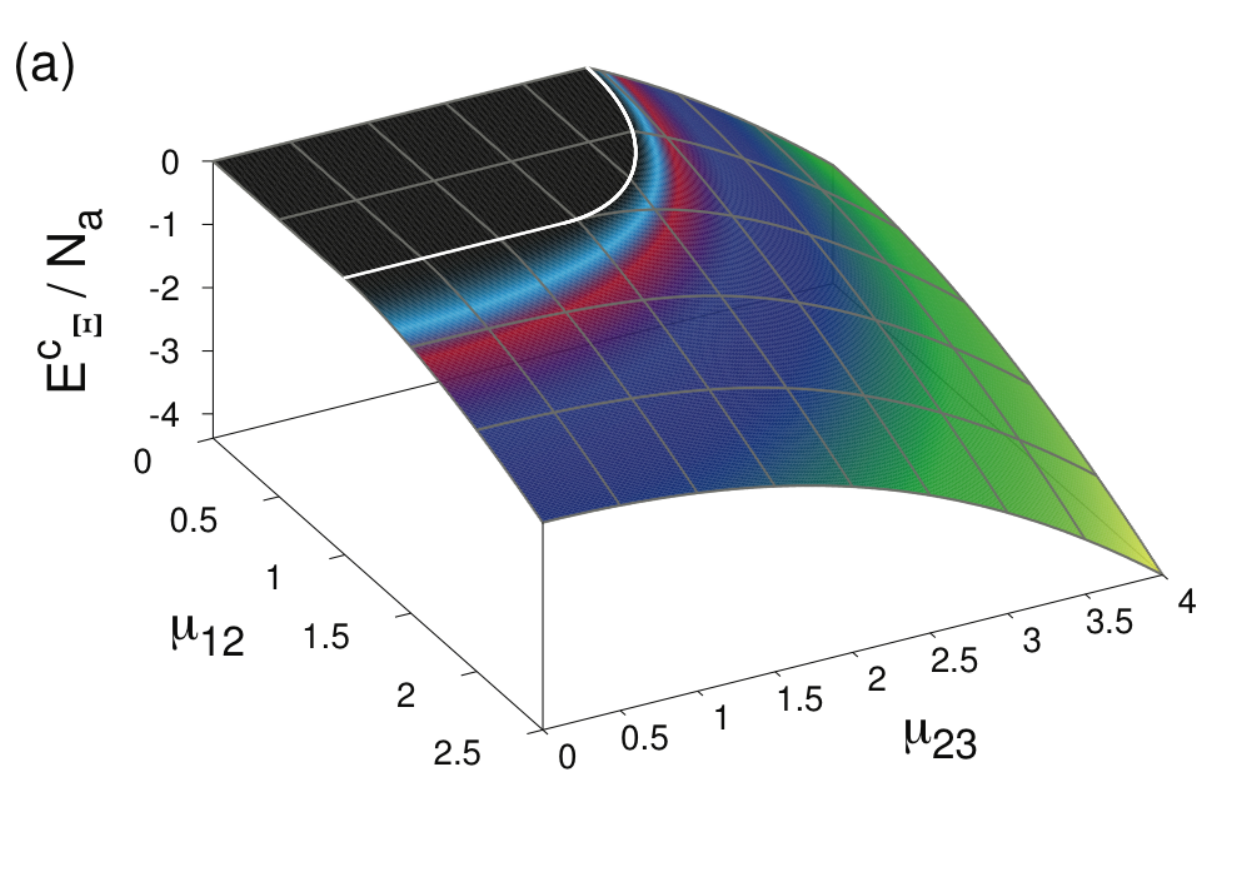}\
\includegraphics[width=0.45\linewidth]{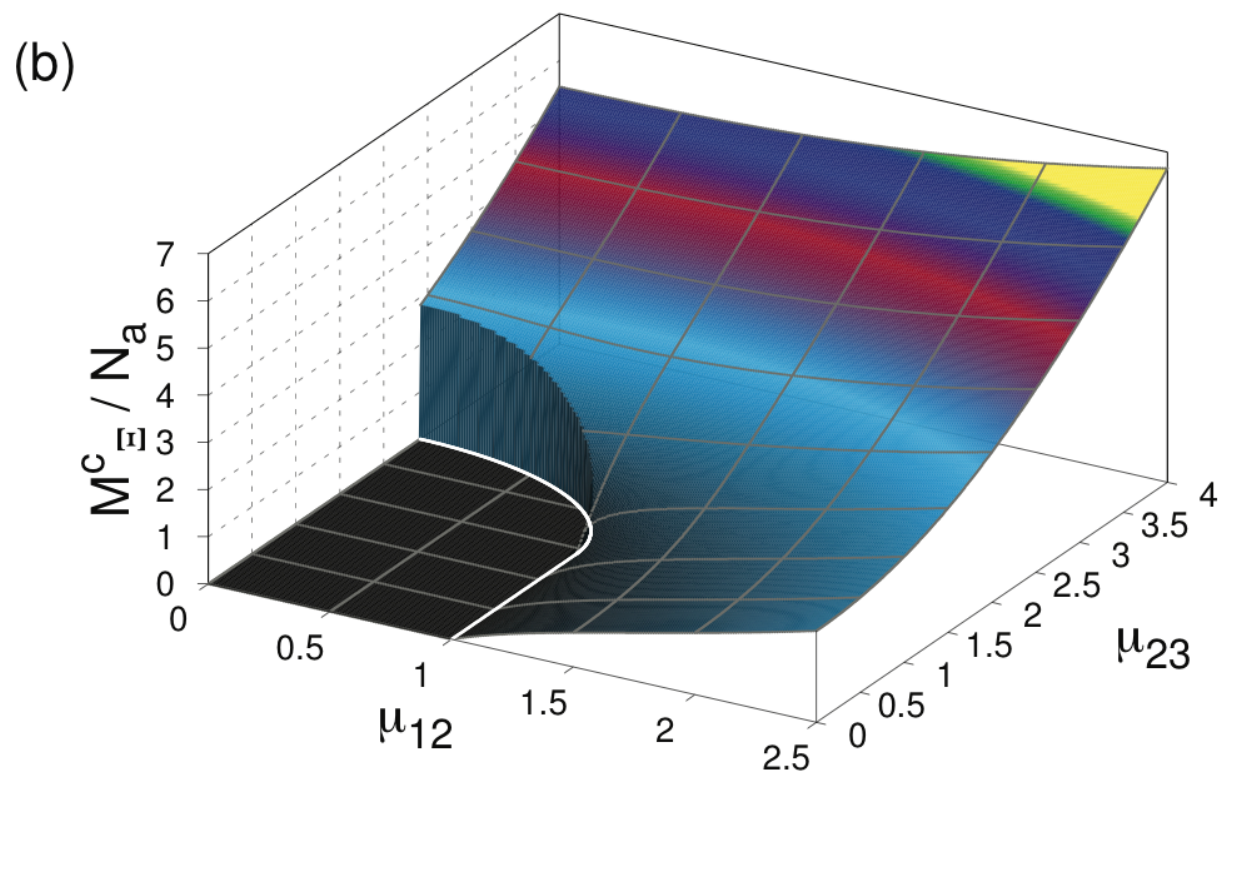} \end{center}
\vspace{-0.5cm}
\caption{
(color online) Numerical results for the lowest value of the energy
surface for the $\Xi$--configuration (a) and its corresponding estimation
for the constant of motion $M_{\Xi}$ (b) , as functions of $\mu_{12}$
and $\mu_{23}$. We consider the resonant case
$\omega_{32}=\omega_{21}=1$,  with $\omega_{1}=0$.}
\label{fig02}
\end{figure}

Using the Ritz variational principle, one finds the best variational
approximation to the ground-state energy of the system and its
corresponding eigenstate for the energy surface~(\ref{eq14}). The
values of $\bar{\varrho}_{c}$ are always given in terms of $\varrho_{2\,c}$ 
and $\varrho_{3\,c}$. These last two, independent of $N_{a}$, are found 
numerically. For the configuration $\Xi$ the estimation of the ground 
energy is plotted in Fig.~\ref{fig02}~(a) as a function of $\mu_{12}$ 
and $\mu_{23}$. In Fig.~\ref{fig02}~(b) we show also the corresponding 
value for the constant of motion $M_{\Xi}$. In both figures we used the resonant case
$\omega_{32}=\omega_{21}=1$, and
$\omega_{1}=0$, with the notation $\omega_{ij}\equiv\omega_{i}-\omega_{j}$.
The white lines indicate the separatrix of the configuration.

\begin{figure} \begin{center}
\includegraphics[width=0.45\linewidth]{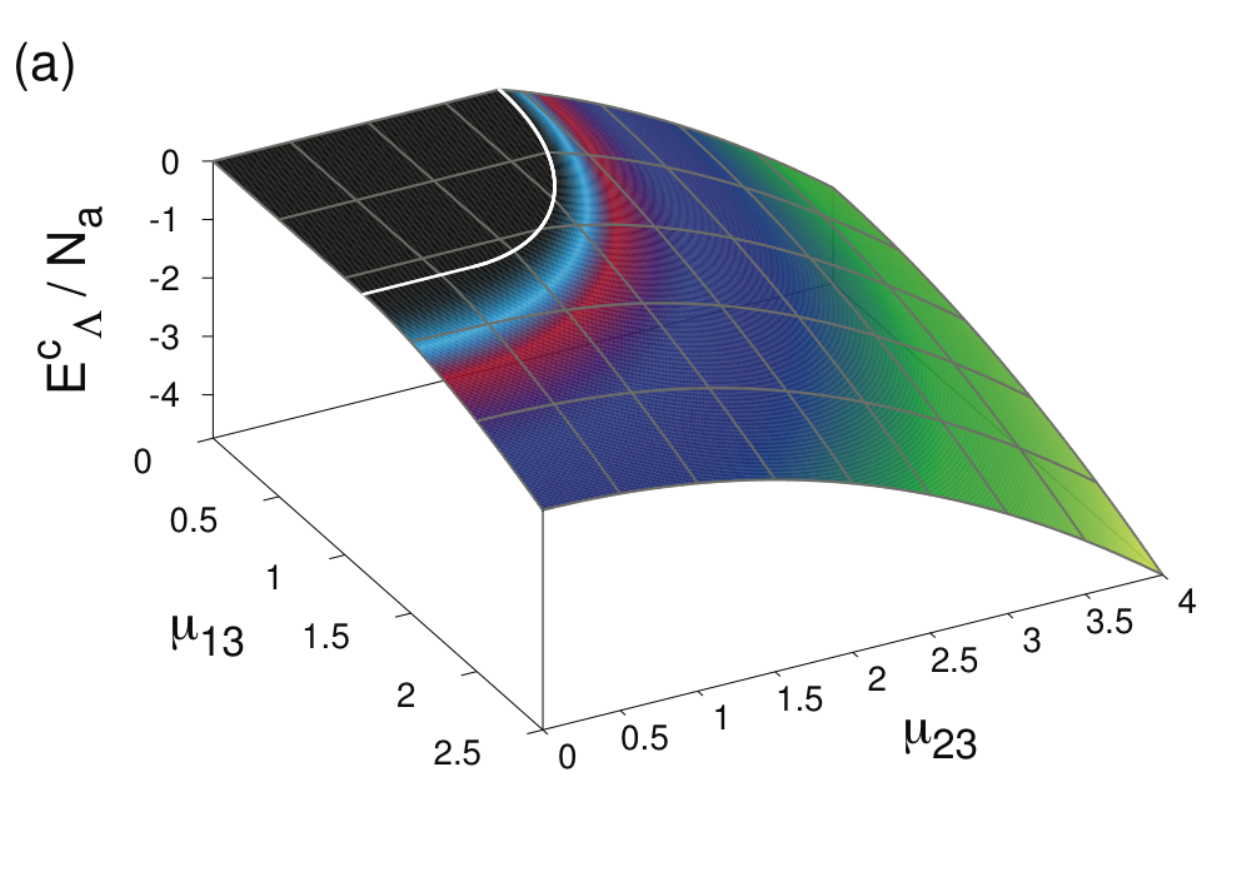}\
\includegraphics[width=0.45\linewidth]{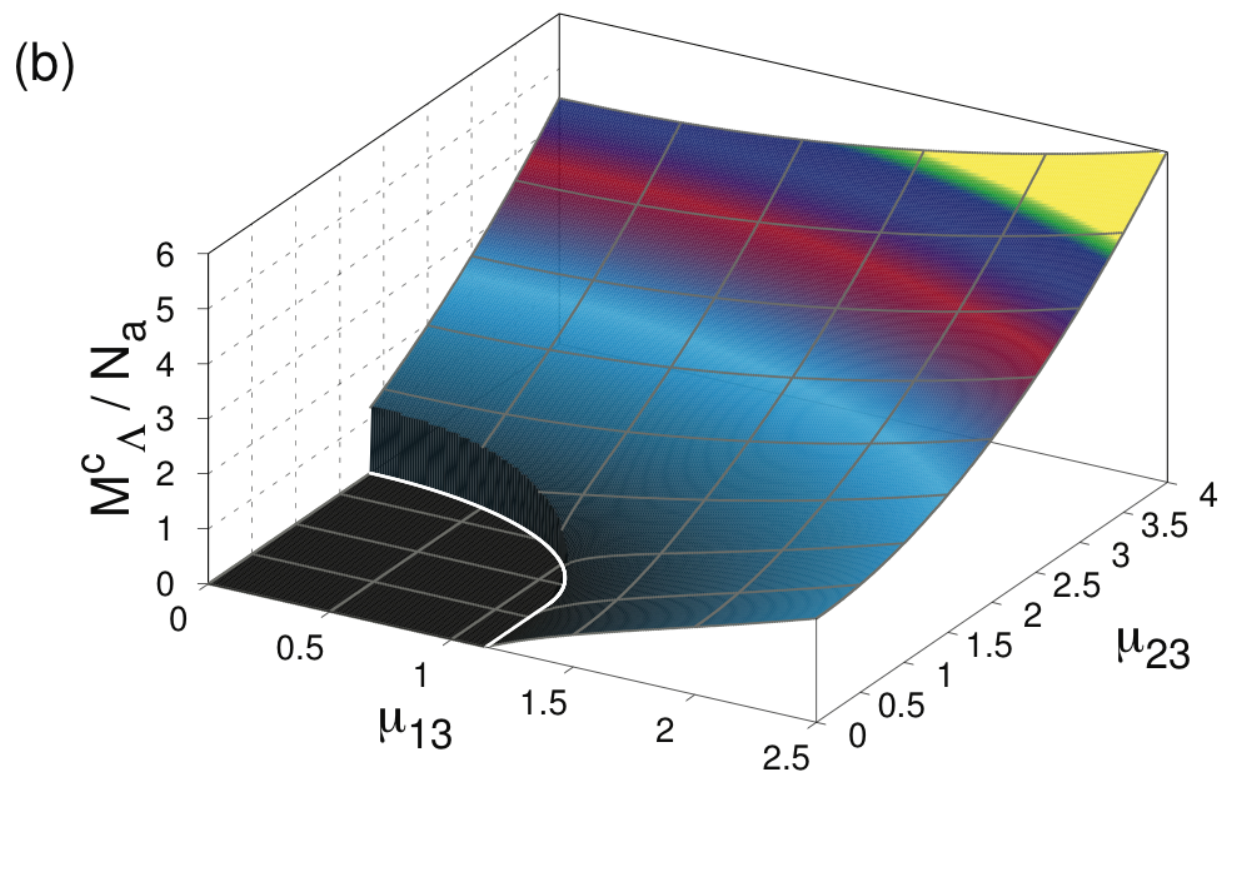} \end{center}
\vspace{-0.5cm}
\caption{
(color online) Numerical results for the lowest value of the energy
surface for the $\Lambda$--configuration (a) and its corresponding
estimation for the constant of motion $M_{\Lambda}$ (b) , as 
functions of $\mu_{13}$ and $\mu_{23}$. We consider the non-resonant case
$\omega_{31}=1.3$,
$\omega_{32}=0.8$, and $\omega_{1}=0$.}
\label{fig03}
\end{figure}

For the configuration $\Lambda$ the result for the semiclassical ground
energy is plotted in Fig.~\ref{fig03}~(a) as a function of $\mu_{13}$ and
$\mu_{23}$. In Fig.~\ref{fig03}~(b) we show the corresponding value of the
constant of motion $M_{\Lambda}$. In both figures we used  $\omega_{31}=1.3$,
$\omega_{32}=0.8$, and $\omega_{1}=0$. Note that, in
this case, we chose for illustrative purposes to work away from 
resonance. Again the white lines denote the separatrix of the system.

% Figure 3

\begin{figure} \begin{center}
\includegraphics[width=0.45\linewidth]{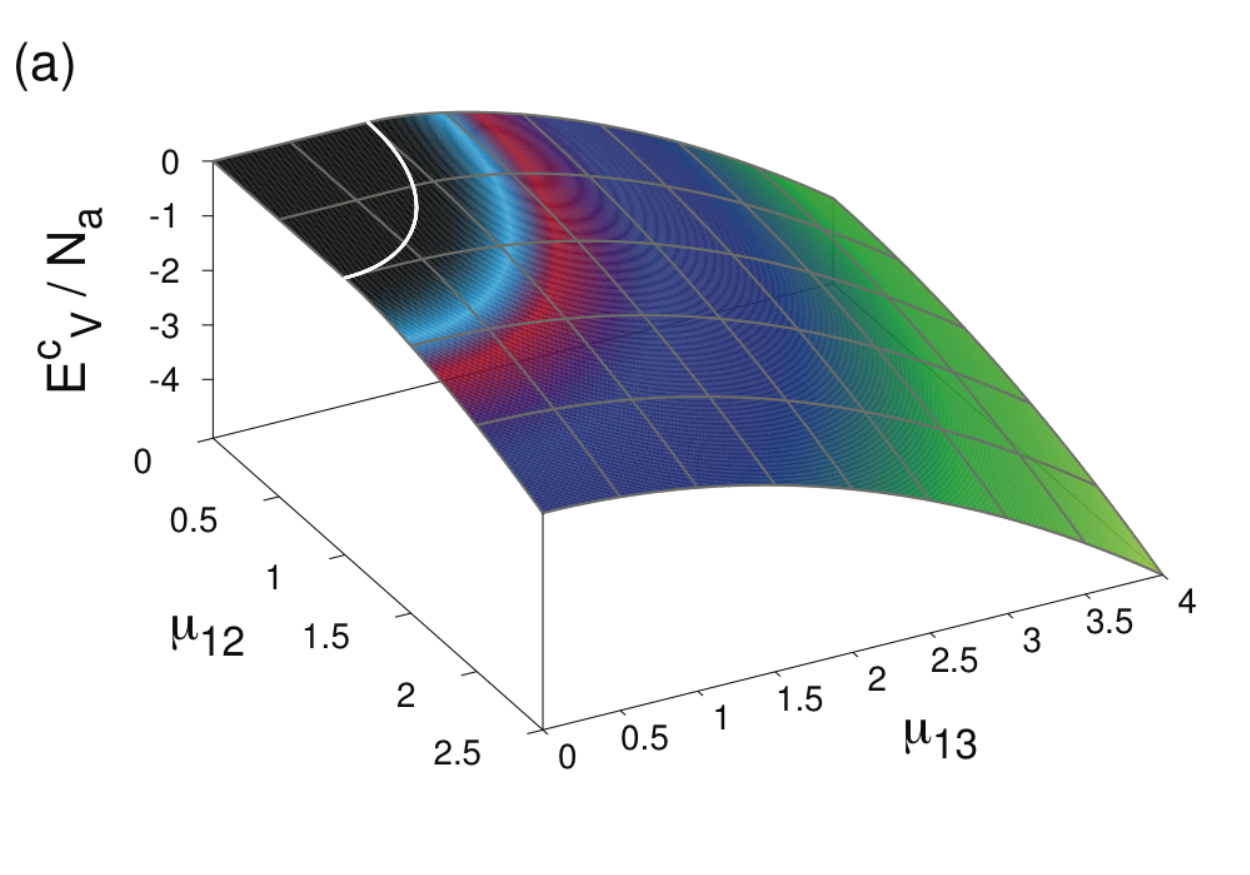}\
\includegraphics[width=0.45\linewidth]{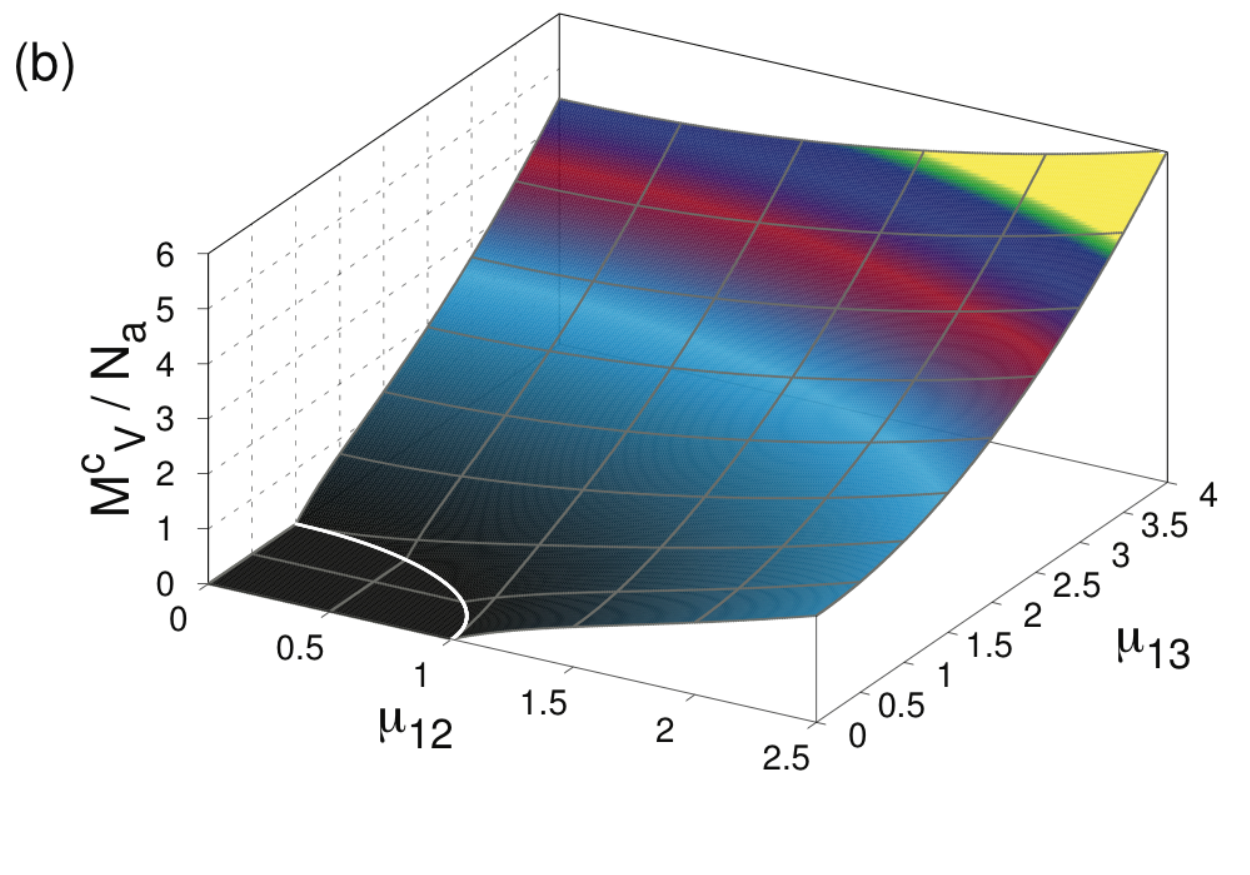} \end{center}
\vspace{-0.5cm}
\caption{
(color online) Numerical results for the lowest value of the energy
surface for the $V$--configuration (a) and its corresponding estimation
for the constant of motion $M_{V}$ (b) , as functions of
$\mu_{12}$ and $\mu_{13}$. We consider  $\omega_{31}=\omega_{21}=1$ with
$\omega_{1}=0$.}
\label{fig04}
\end{figure}

Finally, considering the $V$--configuration we obtain the semiclassical
ground energy shown in Fig.~\ref{fig04}~(a) as a function of
$\mu_{12}$ and $\mu_{13}$, and in Fig.~\ref{fig04}~(b) the
corresponding value of the constant of motion $M_{V}$ is displayed. In
both figures we used again the resonant case
$\omega_{31}=\omega_{21}=1$, with $\omega_{1}=0$. The white line is 
the corresponding separatrix for the configuration.

% Figure 4

The separatrix of the system is obtained by analizing the stability and equilibrium properties of the energy surface 
by means of the catastrophe formalism. Thus, the locus of the points where the
thermodynamic phase transition $M=0 \to M\neq 0$ occurs, is given
in each case by:\\
for the $\Xi$--configuration
    \begin{equation}
	\mu_{12}^{2}+\left[\left|\mu_{23}\right|-\sqrt{\omega_{31}}
	\right]^{2}\,\Theta\left(\left|\mu_{23}\right|-\sqrt{\omega_{31}}
	\right)=\omega_{21}\ ;
	\label{eq15}
    \end{equation}
for the $\Lambda$--configuration
    \begin{equation}
	\mu_{13}^{2}+\left[\left|\mu_{23}\right|-\sqrt{\omega_{21}}
	\right]^{2}\,\Theta\left(\left|\mu_{23}\right|-\sqrt{\omega_{21}}
	\right)=\omega_{31}\ ;
	\label{eq16}
    \end{equation}
and for the $V$--configuration
    \begin{equation}
	\frac{\mu_{12}^{2}}{\omega_{21}}
	+\frac{\mu_{13}^{2}}{\omega_{31}}=1\ .
	\label{eq17}
    \end{equation}
For the $\Xi$--configuration, the phase transition across
$\mu_{12}=\sqrt{\omega_{21}}$ in the separatrix is
of second-order, and the one that takes place along the segment of the
circumference is of first-order. For the $\Lambda$--configuration
something similar happens: across
$\mu_{13}=\sqrt{\omega_{31}}$ in the separatrix the phase
transition is of second-order, and along the segment of circumference
in the separatrix is of first-order.  For the $V$--configuration all
transitions are always of second-order.

4. {\it Quantum case.} In the quantum case we must construct the proper basis for each
configuration taking into account~(\ref{eq07}). Thus, the
basis is characterized by $N_{a}$, the number of atoms, which besides
defines the $U(3)$ highest weight state, and $M$, the total excitation number for
the configuration considered. 

The value of the matrix elements of the $U(3)$ generators in the
Hamiltonian~(\ref{eq01}) in the Gelfand-Tsetlin basis are given in~\cite{moshinsky1967}. 
Using the matrix elements, we calculate the matrix Hamiltonian and through a diagonalization 
procedure obtain the ground state energies, and the expectation value of the total number of 
excitations. In some cases for all the configurations one can get analytic expressions for the 
lowest energy states for a few number of atoms. For example, for the $\Xi$--configuration 
and one atom we obtain
    \begin{equation}
       E_{M_{\Xi}}=M_{\Xi}-\sqrt{M_{\Xi}\,\mu_{12}^{2}
       +(M_{\Xi}-1)\,\mu_{23}^{2}}\ , \nonumber
	\label{eq19}
    \end{equation}
which coincides with the results given in~\cite{yoo&eberly1985} after 
an identification of the parameters used here.

\begin{figure} \begin{center}
\includegraphics[width=0.45\linewidth]{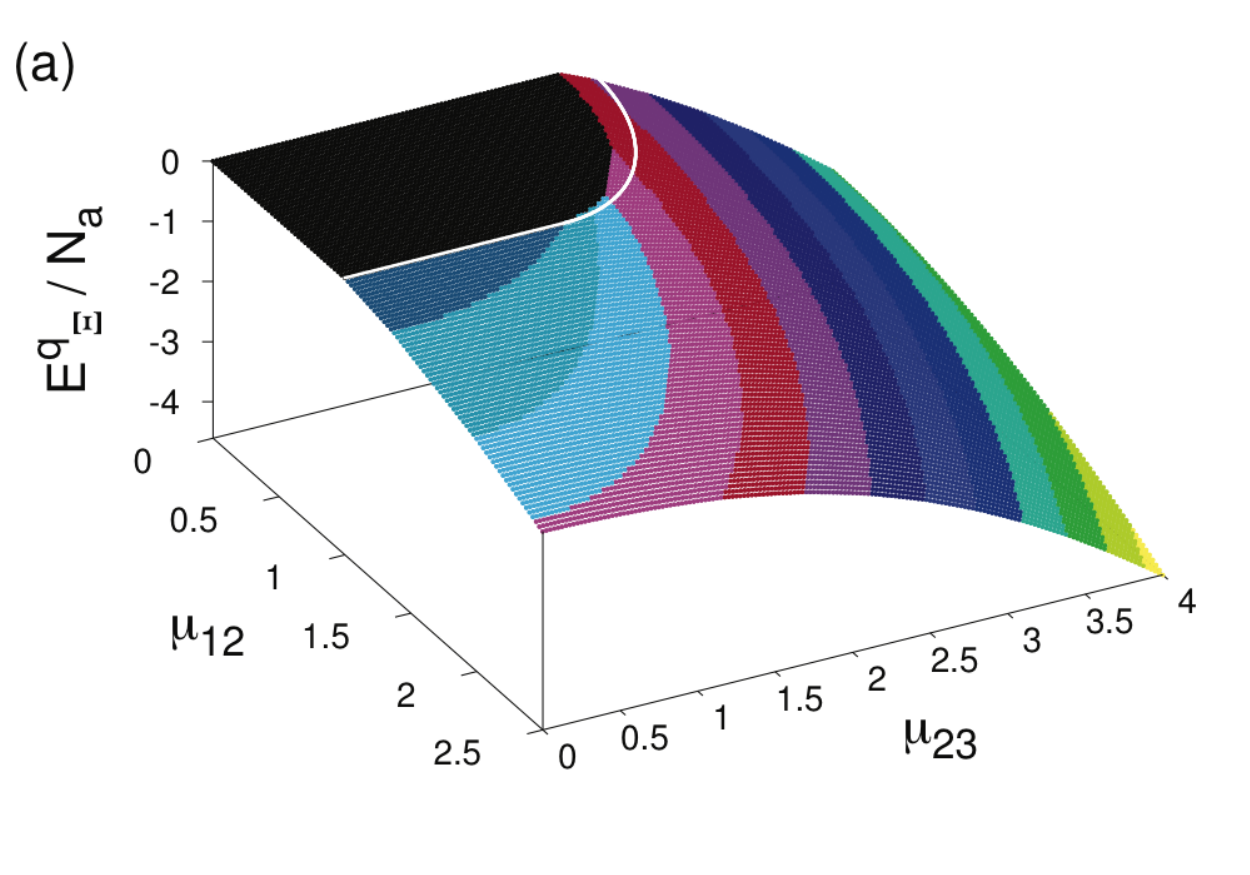}\
\includegraphics[width=0.45\linewidth]{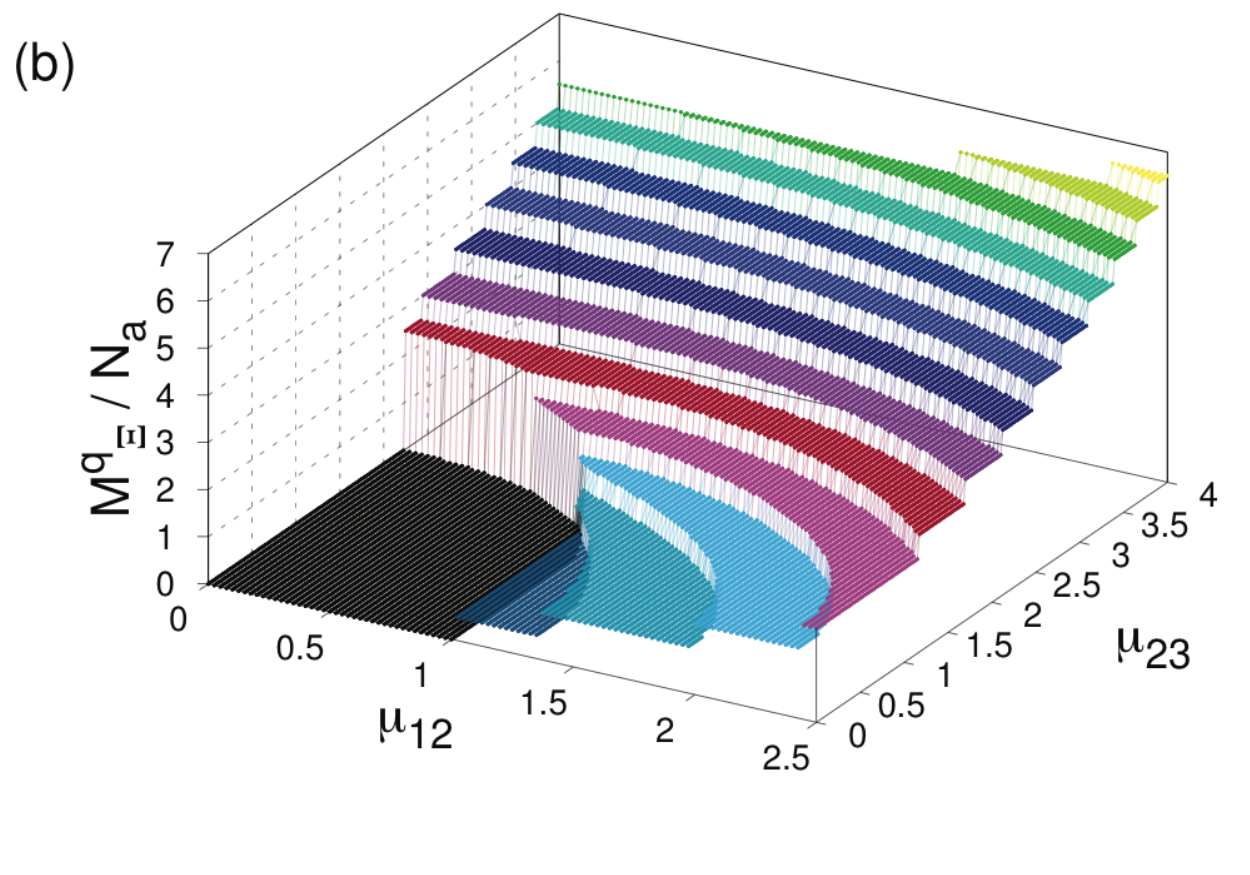} \end{center}
\vspace{-0.5cm}
\caption{
(color online) Numerical results for the ground-state energy of the
model Hamiltonian for the $\Xi$--configuration (a) and the
corresponding value of the constant of motion $M_{\Xi}$ (b), as
functions of $\mu_{12}$ and $\mu_{23}$. We consider the case
$N_{a}=2$, $\omega_{32}=\omega_{21}=1$
and $\omega_{1}=0$. The white line corresponds to the semiclassical
separatrix.}
\label{fig05}
\end{figure}

\begin{figure} \begin{center}
\includegraphics[width=0.45\linewidth]{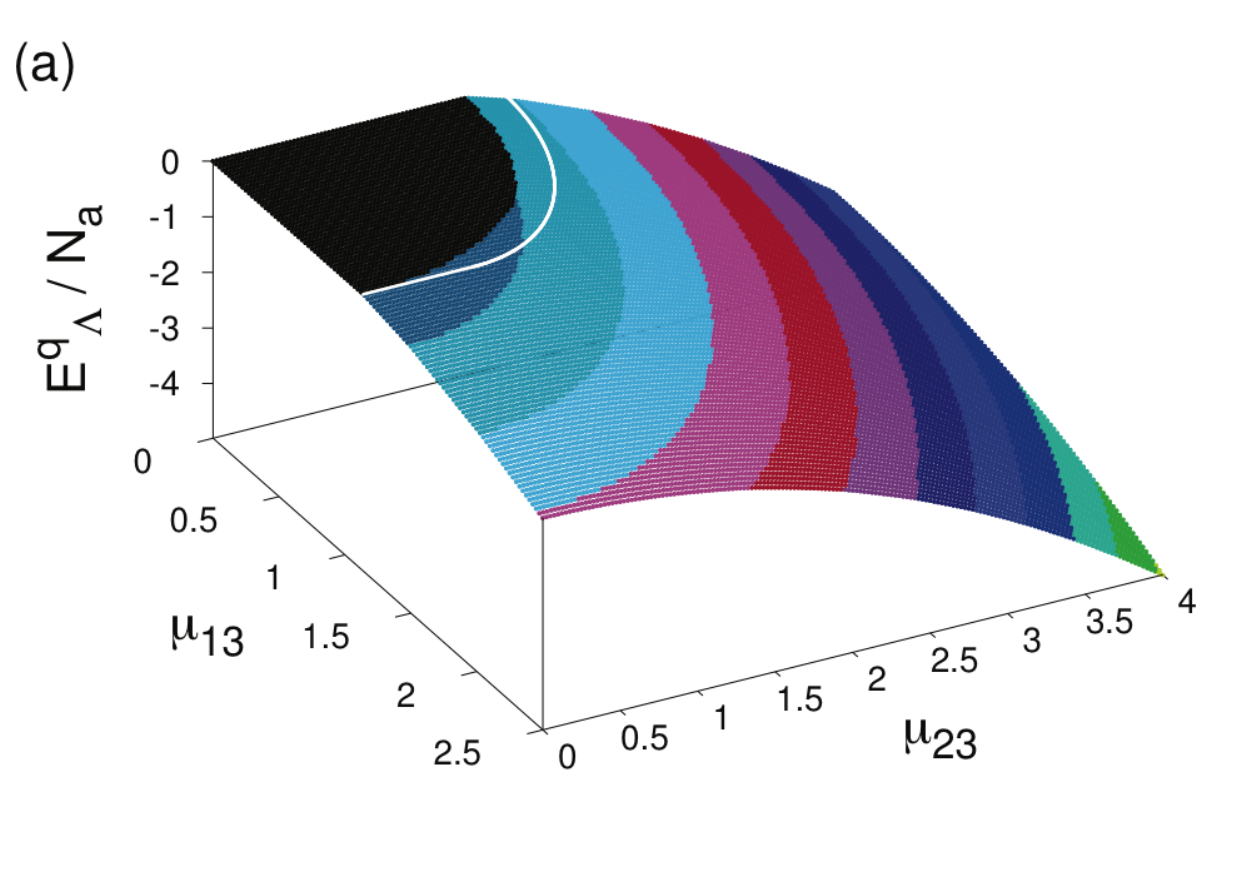}\
\includegraphics[width=0.45\linewidth]{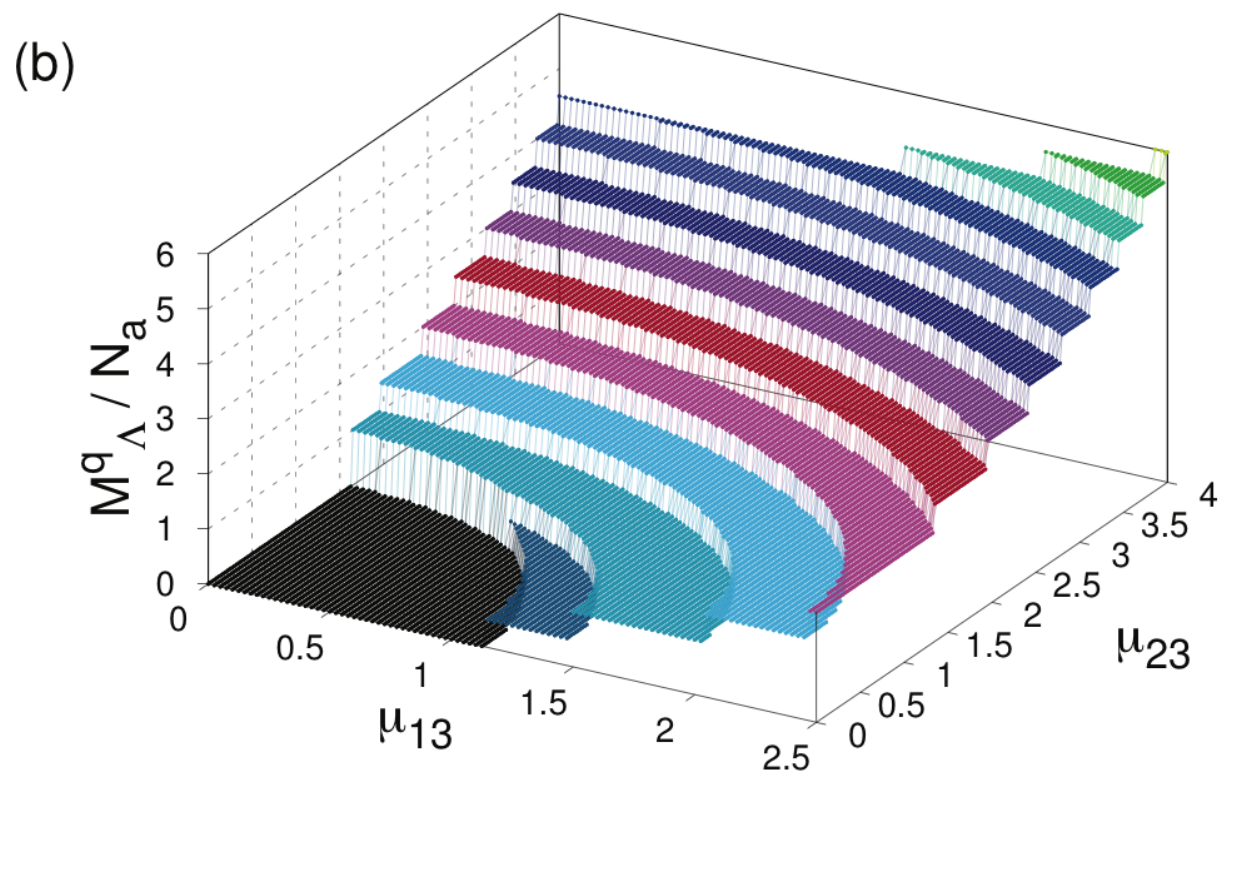} \end{center}
\vspace{-0.5cm}
\caption{
(color online) Numerical results for the ground-state energy of the
model Hamiltonian for the $\Lambda$--configuration (a) and the
corresponding value of the constant of motion $M_{\Lambda}$ (b), as
functions of $\mu_{13}$ and $\mu_{23}$. We consider the case $N_{a}=2$,
$\omega_{31}-\Omega=0.3$,
$\omega_{32}-\Omega=-0.2$ and $\omega_{1}=0$. The white line
corresponds to the semiclassical separatrix.}
\label{fig06}
\end{figure}

\begin{figure} \begin{center}
\includegraphics[width=0.45\linewidth]{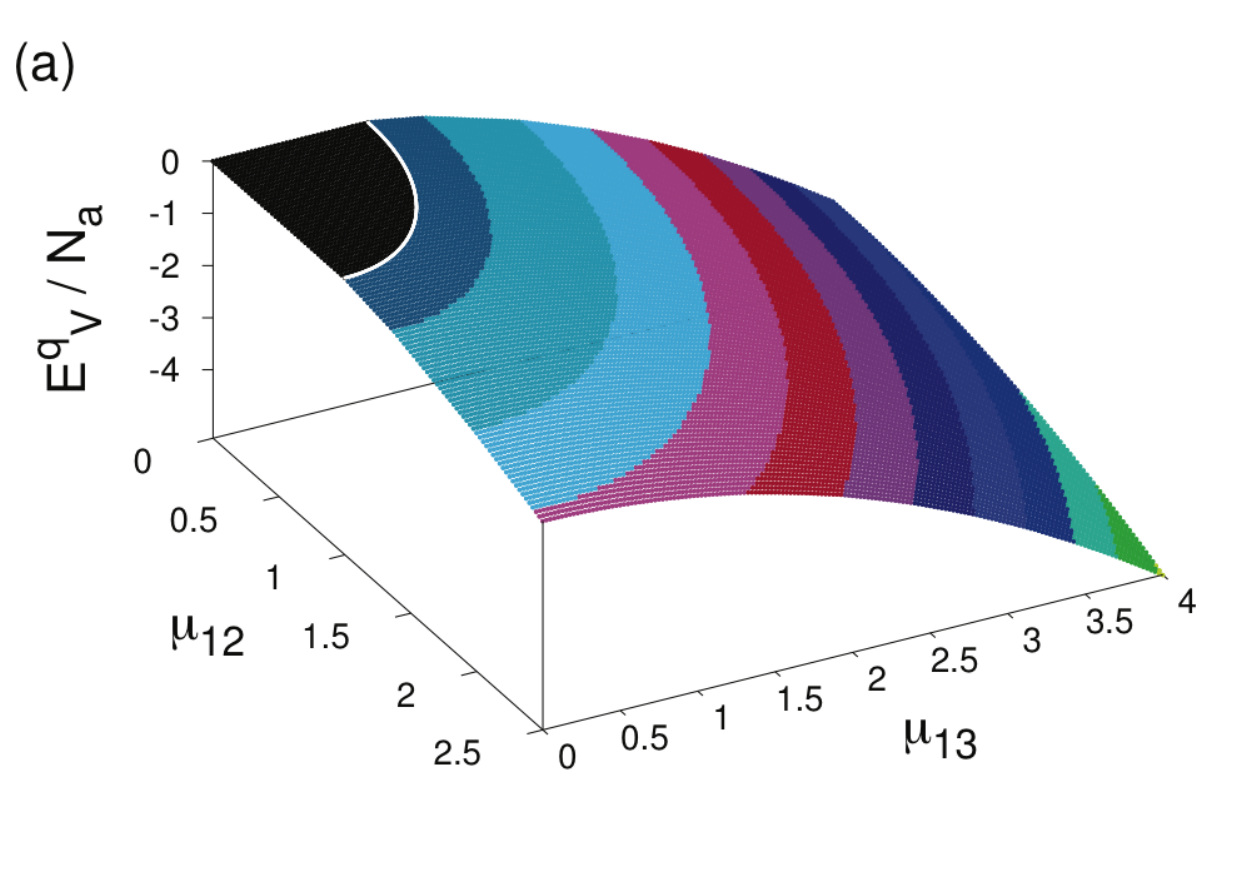}\
\includegraphics[width=0.45\linewidth]{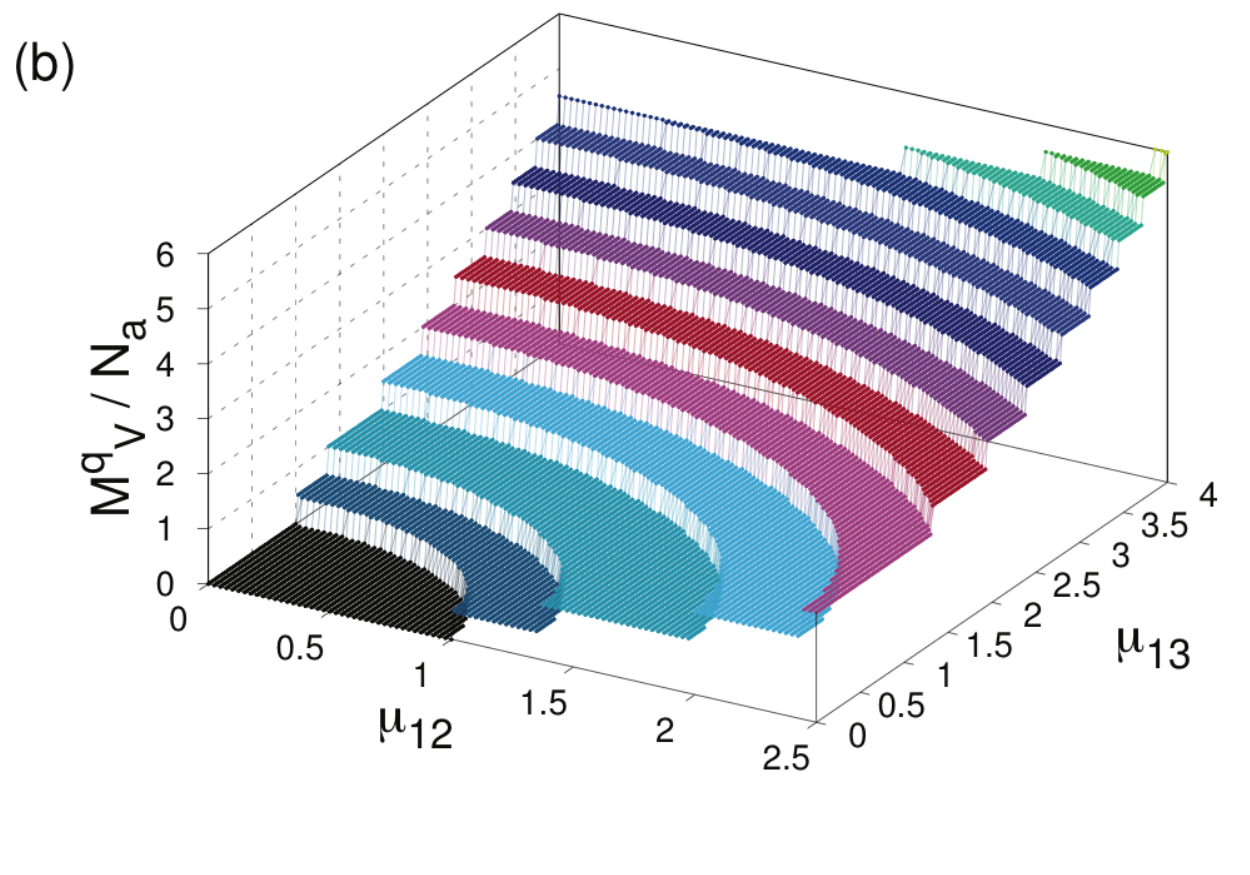} \end{center}
\vspace{-0.5cm}
\caption{
(color online) Numerical results for the ground-state energy of the
model Hamiltonian for the $V$--configuration (a) and the corresponding
value of the constant of motion $M_{V}$ (b), as functions of
$\mu_{12}$ and $\mu_{13}$. We consider the case $N_{a}=2$,
$\omega_{31}=\omega_{21}=1$ and
$\omega_{1}=0$. The white line corresponds to the semiclassical
separatrix.}
\label{fig07}
\end{figure}

In Figs.~(\ref{fig05}), (\ref{fig06}), (\ref{fig07}), we show the
ground-state energy and the value of the constant of motion $M$ for the
$\Xi$--, $\Lambda$--, and $V$--configurations as functions of the
interaction parameters for $N_{a}=2$ atoms, and for the same parameter
values as in the semiclassical case, for comparison purposes. We can
observe that the resemblance between the semiclassical and quantum
results is excellent. In the quantum case we also take notice of phase
crossovers in the ground-state energy which occur every time the
value of the constant of motion changes~\cite{Islam2011}. This may be seen as different 
shades in the energy surface, and is better seen in the plot for
$M$ where there is a jump when a phase crossover takes place.
For the $\Xi$--configuration we see in Fig.~(\ref{fig05}) that there
are phase transitions from the region $M_{\Xi}=0$ directly to regions
$M_{\Xi}=1,\,2,\,\dots,\,5$, for $N_{a}=2$ atoms. When the number of
atoms increases, greater values of $M_{\Xi}$ are pulled towards the
region with $M_{\Xi}=0$, in such a way that there will be phase
transitions from this value to greater $M_{\Xi}$'s. For the other
configurations, one has a similar effect when the number of atoms increase.

% Figure 5

% Figure 6

% Figure 7

\begin{figure} \begin{center}
\includegraphics[width=0.45\linewidth]{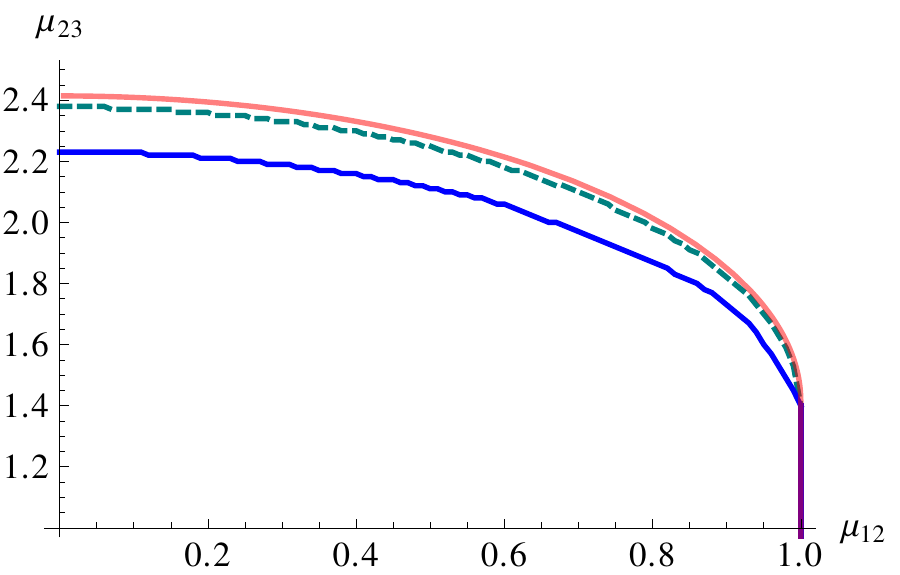}
\end{center}
\vspace{-0.3cm}
\caption{
(color online) Comparison of the equipotential curves of the
ground-state energy of the model Hamiltonian in the
$\Xi$--configuration for $N_{a}=2,\,10$, lower continuous
and middle broken lines respectively, with
$\omega_{32}=\omega_{21}=1$ and
$\omega_{1}=0$. The upper line corresponds to the semiclassical
separatrix, to which all tend as $N_a \to \infty$.}
\label{fig08}
\end{figure}

When the number of atoms $N_{a}$ increases, all the curves where the constant of motion changes 
tend to the classical separatrix. This behavior is shown in Fig.~(\ref{fig08}), where we plot
the equipotential curves for the ground-state energy of the model in
the $\Xi$--configuration with $N_{a}=2,\,10$ atoms for $M_{\Xi}=0$,
and compare with the semiclassical separatrix.

% Figure 8

5. {\it Conclusions.} We studied the three main configurations of a system of $N_{a}$
three-level atoms interacting with a one-mode radiation field, which is
the simplest generalization of the Tavis--Cummings model. Using
$HW(1)$- and $SU(3)$-coherent states, we established the energy
surface for the system and obtained an approximate expression for the
ground-state of the system, as well as the separatrix in the parameters of
the model which defines the locus of the quantum phase transitions. Although
the phase transitions in all the configurations are similar, in the
thermodynamic limit there are only two zones, and the precise form for
the $V$--configuration is qualitatively different to the $\Xi$- and
$\Lambda$--configurations. For the $\Xi$- and
$\Lambda$--configurations the phase transition from one region to the
other can be of first- or second-order, depending on the zone where the
separatrix is crossed; for the $V$--configuration all phase
transitions are of second-order. The comparison with the exact
ground-state of the model shows that our approximation is excellent,
at least for the energy and the expectation value of the constant of
motion of each configuration. We were also able to verify numerically
that when the number of atoms goes to infinity, the multitude of
quantum phase crossovers tend to the thermodynamic limit.

\acknowledgments
This work was partially supported by CONACyT-M\'exico (under project
101541), and DGAPA-UNAM (under projects IN102811, IN102109).


\begin{thebibliography}{999}

\bibitem{tavis-cummings1968}
M. Tavis and F.W. Cummings,
Phys. Rev. {\bf 170}, 379 (1968);
{\bf 188}, 692 (1969).

\bibitem{manko}
{\sl Theory of Nonclassical States of Light}, 
Eds. V.V. Dodonov and V.I. Man'ko (Taylor \& Francis, London, 2003).

\bibitem{baumann2010}
K. Baumann, C. Guerlin, F. Brennecke, and T. Esslinger,
Nature {\bf 464}, 1301 (2010).


\bibitem{nagy2010}
D. Nagy, G. Konya, G. Szirmai, and P. Domokos,
Phys. Rev. Lett {\bf 104}, 130401 (2010).


\bibitem{hepp&lieb1973}
K. Hepp and E. Lieb, Ann. Phys. (N.Y.) {\bf 76}, 360 (1973).\\
Y.K. Wang and F.T. Hioe, Phys. Rev. A {\bf 7}, 831 (1973).

\bibitem{buzek2005}
V. Bu\v{z}ek, M. Orszag, and M. Ro\v{s}ko, Phys.
Rev. Lett. {\bf 94}, 163601 (2005).

\bibitem{castanos2009}
O. Casta\~nos, R. L\'opez-Pe\~na, E. Nahmad-Achar,
J.G. Hirsch, E. L\'opez-Moreno, and J.E. Vitela,
Phys. Scr. {\bf 79}, 065405 (2009).\\
O. Casta\~nos, R. L\'opez-Pe\~na, E. Nahmad-Achar, and J.G. Hirsch,
Phys. Scr. {\bf 80}, 055401 (2009).


\bibitem{li1987}
X.S. Li, D.L. Lin, and C.D. Gong,
Phys. Rev. A {\bf 36}, 5209 (1987);
Z.D. Liu, X.S. Li, and D.L. Lin,
Phys. Rev. A {\bf 36}, 5220 (1987);
D.L. Lin, X.S. Li, and Y.N. Peng,
Phys. Rev. A {\bf 39}, 1933 (1989);
X.S. Li, D.L. Lin, T.F. George, and Z.D. Liu,
Phys. Rev. A {\bf 40}, 228 (1989).


\bibitem{alikskenderov1991}
E.I. Aliskenderov, H.T. Dung, and A.S. Shumovsky,
Quantum Opt. {\bf 3}, 241 (1991);
H.T. Dung, R. Tanas, and A.S. Shumovsky,
Quantum Opt. {\bf 3}, 255 (1991).


\bibitem{yoo&eberly1985}
H.-I. Yoo, and J.H. Eberly,
Phys. Rep. {\bf 118}, 239 (1985).


\bibitem{sung-bowden1979}
C.C. Sung, and C.M. Bowden,
J. Phys. A: Math. Gen. {\bf 12}, 2273 (1979).


\bibitem{brandes2011}
M. Hayn, C. Emary, and T. Brandes,
Phys. Rev. A {\bf 84}, 053856 (2011).


\bibitem{gelfand-tsetlin1950}
I.M. Gelfand and M.L. Tsetlin,
Dokl. Akad. Nauk SSSR {\bf 71}, 825 (1950).


\bibitem{moshinsky1967}
M. Moshinsky, Group Theory and the Many-Body Problem (Gordon and
Breach, New York, 1967).


\bibitem{Islam2011}
R. Islam, et al. Nat. Commun. 2:377 doi:10.1038/ ncomms1374 (2011).

\end{thebibliography}
\end{document}